\documentclass[pra,aps,twocolumn,showpacs,superscriptaddress,nofootinbib]{revtex4}
\usepackage{amsmath,graphicx,epsfig,dsfont}
\def\ket#1{{|#1\rangle}}

\begin{document}

\title{Long-lived heteronuclear spin-singlet states} 

\author{M. Emondts}\email{meike.emondts@rwth-aachen.de}
\address{Institute of Technical and Macromolecular Chemistry, RWTH
Aachen University, Worringer Weg 1, 52074 Aachen, Germany}

\author{M.\ P.\ Ledbetter}\email{micah.ledbetter@gmail.com}
\address{Department of Physics, University of California, Berkeley, California 94720, USA}
\affiliation{AOSense, 767 N. Mary Ave, Sunnyvale, California 94085, USA}

\author{S. Pustelny}
\affiliation{Department of Physics, University of California, Berkeley, California 94720, USA}
\affiliation{Center for Magneto-Optical Research, Institute of Physics, Jagiellonian University, Reymonta 4, PL-30-059 Krak\'ow, Poland}

\author{T. Theis}
\affiliation{Materials Science Division, Lawrence Berkeley National
Laboratory, Berkeley, California 94720, USA}
\affiliation{Department of Chemistry, University of California, Berkeley, California 94720, USA}

\author{B. Patton}
\address{Department of Physics, University of California, Berkeley, California 94720, USA}

\author{J. W. Blanchard}
\affiliation{Materials Science Division, Lawrence Berkeley National
Laboratory, Berkeley, California 94720, USA}
\affiliation{Department of Chemistry, University of California, Berkeley, California 94720, USA}

\author{M. C. Butler}\altaffiliation[Present address:  ]{Environmental Molecular Sciences Laboratory, Pacific Northwest National Laboratory, Richland, Washington 99352, USA}
\affiliation{Materials Science Division, Lawrence Berkeley National
Laboratory, Berkeley, California 94720, USA}
\affiliation{Department of Chemistry, University of California, Berkeley, California 94720, USA}

\author{D. Budker}
\address{Department of Physics, University of California, Berkeley, California 94720, USA}
\address{Nuclear Science Division, Lawrence Berkeley National
Laboratory, Berkeley California 94720, USA}

\author{A. Pines}
\affiliation{Materials Science Division, Lawrence Berkeley National
Laboratory, Berkeley, California 94720, USA}
\affiliation{Department of Chemistry, University of California, Berkeley, California 94720, USA}

\date{\today}


\begin{abstract}
We report observation of long-lived spin-singlet states in a ${\rm ^{13}C-^1H}$ spin pair in zero magnetic field.  In ${\rm ^{13}C}$-labeled formic acid, we observe spin-singlet lifetimes as long as 37 seconds, about a factor of three longer than the $T_1$ lifetime of dipole polarization in the triplet state.  We also observe that the lifetime of the singlet-triplet coherence, $T_2$, is longer than $T_1$. Moreover, we demonstrate that singlet states formed by spins of a heteronucleus and a ${\rm ^{1}H}$ nucleus, can exhibit longer lifetimes than the respective triplet states in systems consisting of more than two nuclear spins. Although long-lived homonuclear spin-singlet states have been extensively studied, this is the first experimental observation of analogous spin-singlets consisting of a heteronucleus and a proton.
\end{abstract}


\pacs{82.56.Hg,33.25.+k,07.55.Jg}

\maketitle

Long-live spin-singlet states have recently attracted considerable attention both experimentally \cite{Caravetta2004,Pileio2010,Sarkar2007, Sarkar2008, Cavadini2005, DeVience2012, Ghosh2011,Pileio2008, Warren2009, Vasos2009, Feng2012} and theoretically \cite{Carravetta2005,Pileio2007,Pileio2009,Pileio2010b,Tayler2011}.  Such states are of practical interest for their ability to store hyperpolarization for more than an order of magnitude longer than $T_1$ \cite{Warren2009,Vasos2009} and for their utility in tracking slow chemical and biological processes \cite{Sarkar2007,Cavadini2005}. For recent reviews of long-lived states in nuclear magnetic resonance (NMR), see Refs. \cite{Levitt2012, Pileio2010b}. 

In the case where two coupled spins are magnetically equivalent, the singlet state is immune to intramolecular dipole-dipole relaxation, often the dominant source of relaxation in NMR.  This is because the dipole-dipole interaction is symmetric with respect to exchange of particles and therefore cannot produce transitions between the antisymmetric singlet and symmetric triplet states. Other sources of relaxation, such as chemical-shift anisotropy, spin-rotation, and effects from paramagnetic impurities are also often suppressed for the singlet compared to the triplet states. As a result, singlet lifetimes can significantly exceed triplet lifetimes. The most dramatic case of a long-lived nuclear spin singlet state is parahydrogen, which has a lifetime of weeks, compared to the $T_1$ lifetime of orthohydrogen, which is on the order of seconds under typical experimental conditions (6 bar hydrogen pressure at room temperature in the absence of paramagnetic impurities).

In high-field NMR such singlet states must be composed of same-species (homonuclear) spin pairs, since the magnetic field breaks the equivalence of spin pairs formed by two different spin species.  Here we demonstrate the existence of heteronuclear spin singlets at zero magnetic field; the spin singlet in question is formed by a strongly coupled $^{13}$C-$^1$H pair in $^{13}$C-labeled formic acid ($^{13}$CHOOH).\footnote{\label{foot:2}Note, the hydroxyl proton can be ignored because 2- and 3-bond couplings between the ${\rm ^{13}C-^1H}$ system and the acidic proton can be neglected due to rapid exchange in protic solvents.} 
We show that the lifetime of this spin singlet can exceed the $T_1$ relaxation time of the triplet-state dipole moment by a factor of three.  Furthermore, the transverse relaxation time of the singlet-triplet coherence exceeds the longitudinal relaxation lifetime of the triplet states, an unusual situation in NMR.  We also show that the extended singlet-state lifetimes are not limited to isolated heteronuclear pairs. In benzene-${\rm ^{13}C_1}$ , with a strongly coupled heteronuclear spin pair weakly coupled to several distant spins, we show that the lifetime of manifolds where the strongly coupled spins are in the singlet state exceeds the lifetime of manifolds where the strongly coupled spins are in the triplet states by about a factor of 1.5. An appealing application for long-lived heteronuclear spin singlets may be in metabolic studies, along the lines of research employing hyperpolarized ${\rm ^{13}C}$-labeled pyruvate \cite{Kohler2007}.

In the present experiments the samples are polarized in a permanent magnet and then pneumatically shuttled to zero field immediately next to the detection cell. DC magnetic field pulses are used to manipulate the polarization in the singlet and triplet manifolds.  This is in contrast to the case of low-field homonuclear singlet spin pairs \cite{Pileio2010}, which require more elaborate multipulse sequences for the interconversion of singlet and triplet states. In the experiments we present here the phase of the resulting signals can be used to identify the type of nuclear spin polarization from which the signal arises.  
The use of a sensitive alkali-vapor magnetometer as a detector allows us to directly probe the resulting spin state. 

The Hamiltonian for two spins $\mathbf{I}$ and $\mathbf{S}$ in the presence of the Zeeman interaction and scalar coupling is 
\begin{equation}
H=J\mathbf{I}\cdot\mathbf{S}+(\gamma_I \mathbf{I}+\gamma_S\mathbf{S})\cdot\mathbf{B}.
\end{equation}
Here, $\gamma_I$ and $\gamma_S$ are the gyromagnetic ratios of the respective spins, and $J$ is the scalar coupling between them. In the high-field limit, eigenstates are those of $I_z$ and $S_z$, $\ket{M_IM_S}$. In zero field, the eigenstates are those of $F^2$ and $F_z$ where $\mathbf{F} = \mathbf{I}+\mathbf{S}$ is the total angular momentum.  We denote the triplet states with $F=1$ as $|T_{+1,0,-1}\rangle$, and the singlet state with $F=0$ as $|S_0\rangle$, where the subscript indicates the magnetic quantum number, $M_F$.  The manifolds are separated in energy by $J$. In zero field, the triplets are symmetric and the singlet is antisymmetric with respect to interchange of the spin labels. Therefore, transitions between the two manifolds due to the symmetric dipole-dipole interaction are forbidden.  As the magnetic field is increased, the $\ket{T_0}$ and $\ket{S_0}$ levels are mixed and dipole-dipole relaxation is gradually turned on.

\begin{figure}
  \includegraphics[width=3.4in]{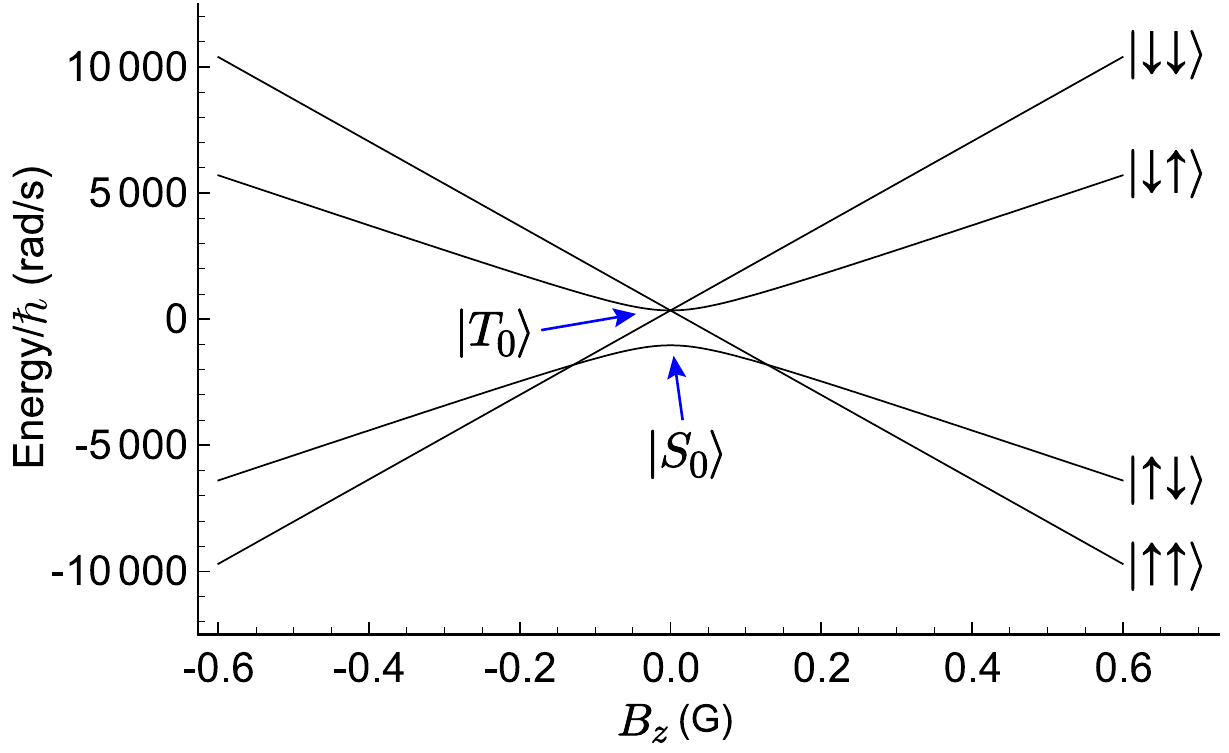}\\
  \caption{Energy levels for two spins, $\bf{I}$ and $\bf{S}$, $I=S=1/2$, in the presence of a scalar coupling and a Zeeman interaction. Here, $J= 220~{\rm Hz}$, and gyromagnetic ratios correspond to carbon and hydrogen.  High-field eigenstates are those of the uncoupled basis, as indicated by the kets on the right.}\label{Fig:energylevels}
\end{figure}

The energy levels of this two-spin system are shown as a function of magnetic field in Fig. \ref{Fig:energylevels}.
Polarization of the zero-field levels is achieved in our experiment by transferring the sample from thermal equilibrium in a prepolarizing magnet (20 kG) to zero field. The rate of transit (occurring over 1.5 s) is slow compared to $J$, populations of the thermally polarized high-field states are adiabatically transferred to those of the low-field states.
This results in an excess population of the $\ket{S_0}$ state over the $\ket{T_0}$ state, and an excess of $\ket{T_{+1}}$ state over $\ket{T_{-1}}$.  The latter corresponds to the dipole moment of the triplet state oriented in the direction of the quantization axis ($z$).  More details of polarization via adiabatic transfer can be found in the Supplementary Material. After adiabatic transfer from high to zero field, the sample stays in zero field for a certain storage time $\tau$. DC magnetic-field pulses are used to excite a coherence between zero-field substates and to manipulate the polarization in the singlet and triplet manifolds. The use of a sensitive alkali-vapor magnetometer (described in \cite{Ledbetter2011}) allows us to directly probe the resulting spin state. We observe the $z$ component of the nuclear magnetization, proportional to the trace of $\rho(\gamma_I I_z+\gamma_S S_z)$.  Since the term in parenthesis is a vector operator with magnetic quantum number equal to zero, observable transitions are those with $\Delta F = \pm 1$ and $\Delta M_F = 0$. These are transitions between the $\ket{T_0}$ and $\ket{S_0}$ states, which produce magnetization oscillating in the $z$ direction with frequency~$J$.

Coherences between $\ket{T_0}$ and $\ket{S_0}$ can be established from excess population in the singlet state by application of a DC pulse of magnetic field in the $z$ direction, resulting in $z$ magnetization proportional to $\sin(2 \pi J t)$. This corresponds to a dispersive peak in the real part of the Fourier transform of the magnetometer signal.  The amplitude of the signal in this case is proportional to the difference in population between the $\ket{T_0}$ and $\ket{S_0}$ states.  Coherences between $\ket{T_0}$ and $\ket{S_0}$ can also be established from dipole polarization in the triplet state by application of a pulse of magnetic field in the $x$ direction, resulting in $z$ magnetization proportional to $\cos( 2\pi J t)$, corresponding to an absorptive peak in the real part of the Fourier transform. In this case, the amplitude is proportional to the difference in population between the $\ket{T_{+1}}$ and $\ket{T_{-1}}$ states. Details of how coherences are produced from the different polarization moments and the resulting signals are presented in the Supplementary Material.

Figure \ref{Fig:signals} shows the real part of the Fourier transform of the magnetometer signal following adiabatic transition of the formic acid sample (FAIII) to zero field and subsequent application of magnetic-field pulses in either $x$ or $z$ direction.  The phase is in agreement with the discussion above, with absorptive and dispersive lineshapes for $x$ and $z$ pulses, respectively. Overlaying each trace is the real part of a fit to a complex Lorentzian, with half-width at half-max (HWHM) linewidth equal to about 10 mHz, corresponding to coherence lifetime $T_2 = 16~{\rm s}$. In both cases there is a small deviation from Lorentzian line shape.  We attribute this to variations in $J$ across the sample due to temperature gradients, as discussed further in the Supplementary Material.

\begin{figure}
  \includegraphics[width=3.4in]{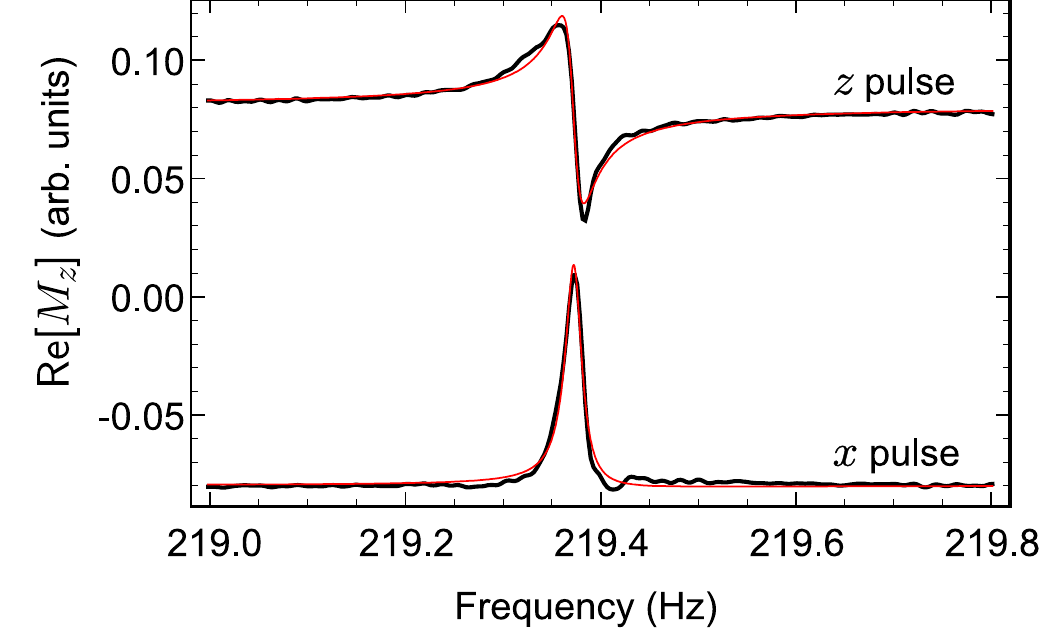}\\
  \caption{Zero-field NMR signals of formic acid (sample FAIII).  The real part of the signal following a pulse in the $z$ direction originates from an initial population imbalance between the $\ket{S_0}$ and $\ket{T_0}$ states and is dispersive.  The real part of the signal due to a pulse in the $x$ direction originates from dipole moment in the triplet manifold (excess of $\ket{T_{+1}}$ over $\ket{T_{-1}}$ states), and is absorptive.
  }\label{Fig:signals}
\end{figure}

\begin{figure}
  \includegraphics[width=3.4in]{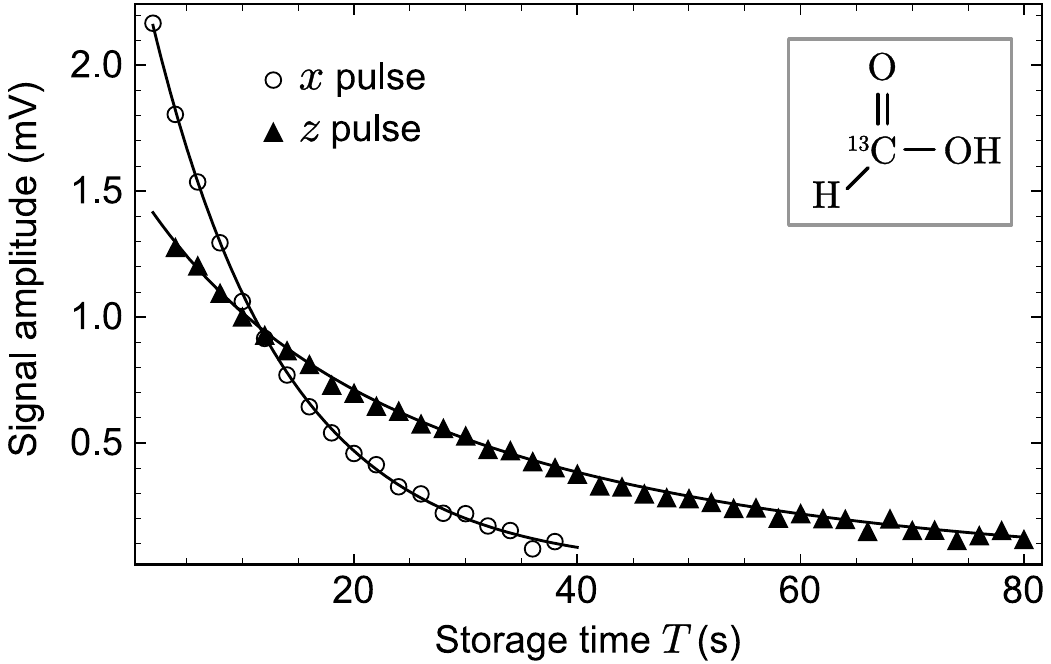}\\
  \caption{Decay of signal amplitude as a function of storage time for magnetic-field pulses applied in the indicated directions using sample FAIII.  The solid lines overlaying the data are fits to single (double) exponentials for \textit{x (z)} magnetic-field pulses.
  }\label{Fig:signaldecay}
\end{figure}

The amplitude of the zero-field NMR signal as a function of the storage time $\tau$ for application of magnetic field pulses in the $x$ or $z$ directions is shown in Fig. \ref{Fig:signaldecay} for sample FAIII. The decay of the signal amplitude obtained by application of $x$ pulses is well described by a single decaying exponential, with time constant $T_1 = 11.8(1)~{\rm s}$.  We note that our definition of $T_1$ deviates slightly from the high-field case, which usually corresponds to the lifetime of population difference between single-particle Zeeman eigenstates.  Here, $T_1$ is the lifetime of the difference in population between the triplet states $\ket{T_{+1}}$ and $\ket{T_{-1}}$ in the coupled system.  The signal obtained by application of a $z$ pulse is well described by a double exponential with fast and slow time constants $T_{f} = 10(3)$ and $T_{s} = 37(2)$ s.  The slow decay is due to transitions between the singlet and triplet states and reflects the lifetime of the singlet state. The fast decay is a result of equilibration within the triplet states, which reduces the population imbalance between the $\ket{T_0}$ and $\ket{S_0}$ states.  Single and biexponential decays for relaxation of dipole and singlet polarization are predicted by a phenomonological rate-equation model, described in the Supplementary Material.  

Formic acid samples were mixtures of ${\rm ^{13}C}$-labeled formic acid, acetonitrile, and ${\rm H_2O}$. The samples were flame sealed in 5-mm NMR tubes after four or five cycles of freezing and thawing under vacuum in order to remove dissolved gases, in particular oxygen, which can cause additional relaxation. 

The main results for three different samples (containing formic acid, acetonitrile and water) are presented in Table \ref{Tab:decays}. Dilution of formic acid with acetonitrile/water mixtures was performed to reduce intermolecular interactions and to promote fast exchange of the hydroxyl proton. As can be seen in Table \ref{Tab:decays}, an increased fractional content of acetonitrile increases the singlet lifetime, likely due to a reduction in viscosity and decreased correlation times. It is interesting to note that the lifetime of the singlet-triplet coherence is longer than the $T_1$ relaxation time for all samples listed in Table \ref{Tab:decays}.  This is a situation rarely encountered in high-field NMR because $T_1$ relaxation in high field corresponds to exchange of energy between two Zeeman eigenstates, and $T_2$ relaxation corresponds to dephasing of a coherence between the same two Zeeman eigenstates.  In the zero field case at hand, the $T_1$ relaxation of the dipole moment corresponds to relaxation of population differences between the states $\ket{T_{\pm 1}}$, whereas the observed singlet-triplet coherence involves the states $\ket{T_0}$ and $\ket{S_0}$.

\begin{table}
  \centering
  \begin{tabular}{|c|c|c|c|c|}
    \hline
    Sample and content (FA,A,${\rm H_20}$) & $T_1 ~{\rm(s)}$ & $T_f~{\rm (s)}$ & $T_s ~{\rm(s)}$ & $T_2$\\
    \hline
    FAI (100,0,10) & 5.8(1) & 2.4(3) & 18.4(3) & 8.0(5) \\
    FAII (50,100,10) & 8.0(5) & 8.84(7) & 26.5(5) & 10.3(3)\\
    FAIII (5,50,5) & 11.8(1) & 10(3) & 37(2) & 16.2(4)\\
    \hline
  \end{tabular}
  \caption{Summary of contents (in ${\rm \mu L}$) and decay times $T_1$, $T_f$, $T_s$ and $T_2$ for several samples. FA = formic acid, A = acetonitrile.}\label{Tab:decays}
\end{table}

The theory of nuclear spin relaxation in two-spin systems has been studied extensively for the case of homonuclear spins in low magnetic field \cite{Carravetta2005,Pileio2009,Tayler2011,Pileio2010b} and can be found in Ref. \cite{Carravetta2005}. These formulas can be adapted to heteronuclear systems. Rate expressions for relaxation due to dipole-dipole interactions or external fluctuating fields can be found in  Ref. \cite{Carravetta2005}.  In the case of dipole-dipole relaxation,, the ratio of $T_2$ to $T_1$ should be $T_2/T_1 = 9/5$. If externally fluctuating fields due to paramagnetic impurities or other nuclei (for example, the hydroxyl proton) limit the lifetime, we expect a ratio $T_2/T_1 = 2$, assuming that the fields are uncorrelated at the two nuclei. 
We measure $T_2$/$T_1$ to be 1.28 -- 1.37, which indicates that $T_2$ may be limited by temperature or magnetic-field inhomogeneities.

We suspect that exchange of the hydroxyl proton, or possibly some paramagnetic impurity, ultimately limits the lifetime of the singlet state. This is supported by the observation that all lifetimes have a roughly linear dependence on temperature.  In sample FAII, $T_s$ decreases by about 10\% as the temperature is increased in the range of $20-60~{\rm ^\circ C}$, while $T_f$, the fast  component of the relaxation, increases by a similar fractional amount. We observe that $T_1$ increases by about 10\% over this range, consistent with the expectation that dipole-dipole relaxation is suppressed by improved motional narrowing at elevated temperatures. 

\begin{figure}
  \includegraphics[width=3.3 in]{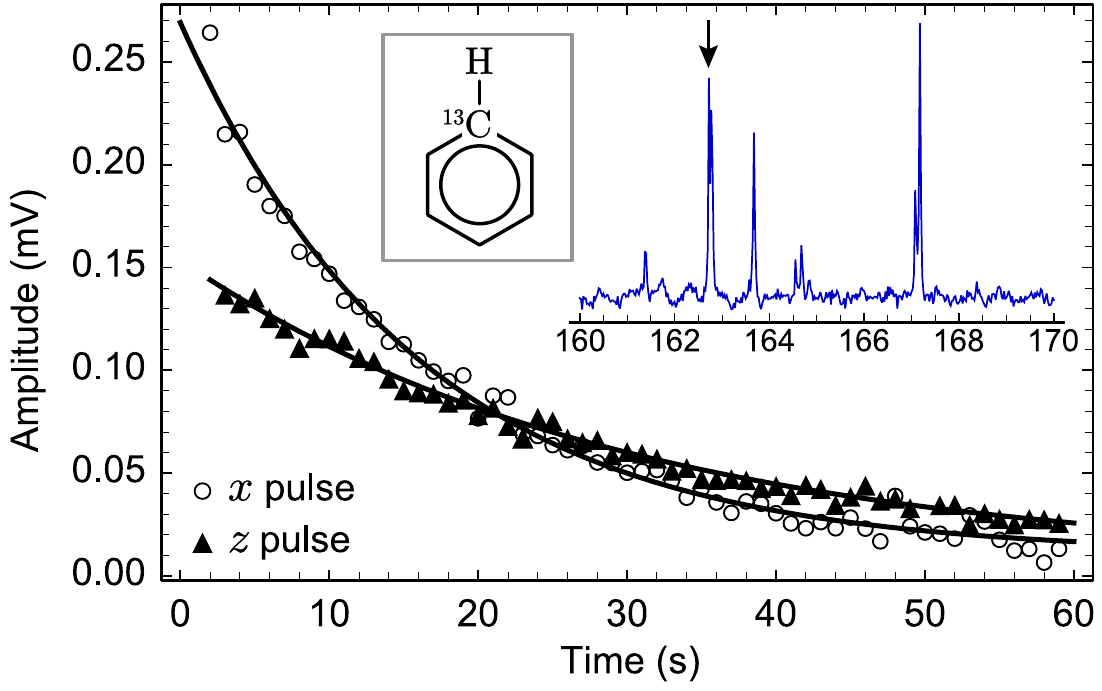}\\
  \caption{Decay of signal amplitude in benzene-${\rm ^{13}C_1}$  for magnetic field pulses applied in the $x$ (circles) and $z$ (triangles) directions (this spectrum is the result of averaging five transients). The solid lines overlaying the data are fits to single decaying exponentials.  The inset shows a portion of the zero-field spectrum, and the arrow indicates the peak from which these decay curves are extracted.}\label{Fig:benzdecay}
\end{figure}

The increased lifetime of the singlet state is not strictly limited to the case of two-spin systems \cite{Pileio2006,Pileio2007,Ahuja2009}.  In systems where there are two strongly coupled spins and a series of weaker couplings, we also observe an extension in the lifetime of the manifolds where the strongly coupled system is in the spin-singlet state.  Singly labeled ${\rm ^{13}C}$-benzene is one such example, in which there is a strongly coupled ${\rm ^1H-^{13}C}$ system, weakly coupled to a set of distant spins, resulting in splitting of the single zero-field NMR line.  A portion of the zero-field spectrum of singly labeled ${\rm ^{13}C}$ benzene in the neighborhood of the one-bond J-coupling frequency, acquired after application of an $x$ directed pulse, is shown inset in Fig. \ref{Fig:benzdecay}. A complete discussion of the zero-field spectrum resulting from ${\rm ^{13}C}$-labeled benzene is presented in Ref. \cite{Blanchard2013}.  Individual lines are typically about 10~mHz (HWHM).  The signal is absorptive as expected from dipole polarization of the triplet manifolds.  Application of a pulse in the $z$ direction produces a signal with the same frequencies and with dispersive lineshape, as expected from polarization of the singlet state. The decay of the signal amplitude (for the peak indicated by the arrow in the inset) is given in the main panel for $x$ pulses (corresponding to triplet polarization) and $z$ pulses (corresponding to singlet polarization). While multiple decay rates are likely, our signal-to-noise ratio is not sufficient to cleanly differentiate between them, so we fit these data to single decaying exponentials, as indicated by the solid lines.  The decay times extracted from the fit are $T_1 =19(1)~{\rm s}$ and $T_s = 30(2)~{\rm s}$, for pulses applied in the $x$ and $z$ directions, respectively.  Decay curves for the other peaks yield similar lifetimes.  These results confirm that the long lifetime of heteronuclear spin singlet states is not limited to the case of two isolated heteronuclear spins, but can be extended to systems consisting of two strongly coupled heteronuclear spins with weakly coupled distant spins.

In conclusion, we have shown that the lifetime of heteronuclear spin-singlet states in zero magnetic field can be substantially longer than that of the dipole moment in the triplet state. In ${\rm ^{13}C}$-labeled formic acid, we find that the singlet lifetime is a factor of three longer than the lifetime of the triplet dipole moment. The extended lifetime of singlet states in homonuclear spin pairs has been known for some time, but extended lifetime of heteronuclear spin singlets has not been observed before. One advantage of working with heteronuclear spin-singlets is that they can be manipulated by DC magnetic-field pulses to produce observable magnetization.  The use of sensitive atomic magnetometers, as employed here, enables direct observation of this magnetization in zero field.
Typically, singlet states are not eigenstates of the high-field Zeeman Hamiltonian, and if a high degree of symmetry is not present in the molecule of choice, RF spin locking has to be implemented in order to remove chemical shifts. At zero magnetic field, heteronuclear as well as homonuclear spin systems are naturally coupled into long-lived eigenstates.  This vastly expands the range of chemical systems which can exhibit long-lived spin order, motivating further research in zero-field NMR as a valuable spectroscopic tool.

\textit{Acknowledgements}: This research was supported by the U.S. Department of Energy, Office of Basic Energy Sciences, Division of Materials Sciences and Engineering under Contract No. DE-AC02-05CH11231 (JWB, TT, and AP), by the National Science Foundation under award CHE- 0957655 (DB, MCB and MPL), and by the Kolumb program of the Foundation for Polish Science (SP). JWB is also supported by a National Science Foundation Graduate Research Fellowship under Grant No. DGE-1106400. MPL appreciates useful discussions with B. Koelsch. We are grateful to S. Appelt for stimulating discussions and support.

\end{document}